\documentclass[aps,prl,twocolumn,amsmath,reprint,superscriptaddress,longbibliography]{revtex4-2} 
\usepackage{hyperref}
\usepackage{graphicx} 
\usepackage{units}
\usepackage{upgreek}
\usepackage{color}
\usepackage{amsmath,amssymb,amsfonts}
\usepackage[dvipsnames]{xcolor}

\newcommand{\vv}{{\bf v}}

\newcommand{\hn}{{{{\bf n}}}}

\newcommand{\dens}{\rho}

\begin{document}

\title{Active thermodynamics of inertial chiral active gases: equation of state and edge currents}

\author{Lorenzo Caprini}
\email{lorenzo.caprini@uniroma1.it} 
\affiliation{Sapienza University of Rome, Piazzale Aldo Moro, 5, 00185, Rome, Italy}

\author{Umberto Marini Bettolo Marconi}
\affiliation{University of Camerino, via Madonna delle Carceri, 62032, Camerino, Italy}

\author{Benno Liebchen}
\affiliation{Technische Universit\"at Darmstadt, Institute of Condensed Matter Physics, Hochschulstrasse 8, 64289 Darmstadt, Germany}

\author{Hartmut L\"owen}
\affiliation{Heinrich-Heine-Universit\"at D\"usseldorf, D-40225 D\"usseldorf, Germany}

\date{\today}

\begin{abstract}  
One of the most fundamental quests in the physics of active matter concerns the existence of a comprehensive theory for its macroscopic properties, i.e.\ an ``active thermodynamics". Here, we derive and experimentally verify key elements of the active thermodynamics of ideal chiral active gases, unveiling edge currents and odd diffusivity as their peculiar features. Our main results are the derivation of an equation of state relating density and pressure via a chirality-dependent effective temperature, the derivation of Fick's law including the full diffusion matrix predicting odd diffusion, and the exact prediction of edge currents at container walls that nonmonotonically depend on chirality.
\end{abstract}

\maketitle

\paragraph{Introduction --}

Equilibrium thermodynamics of gases is anchored by the renowned ideal gas law -- an equation of state that connects pressure and density through environmental temperature. Picture an equilibrium gas as a fluid composed of inert or passive particles at low density, moving due to thermal agitation. These systems are described by the Maxwell-Boltzmann distribution and vanishing currents due to thermal equilibrium with the environment.

However, in the realms of biology and soft matter, several systems maintain permanent currents that keep them far from equilibrium.
A key example is active matter~\cite{marchetti2013hydrodynamics, elgeti2015physics} comprising self-propelled particles, including bacteria, algae or synthetic microswimmers that use energy from their environment to propel themselves \cite{bechinger2016active}.
There has recently been an enormous interest in exploring the thermodynamic properties of active systems. Specifically, it is now well known that the large-scale behavior of the active ideal gas can be mapped onto an equilibrium system with an enhanced temperature~\cite{palacci2010sedimentation, szamel2014self, petrelli2020effective, hecht2024define}. Thus, even if far from equilibrium locally, the thermodynamic properties of the active ideal gas do not feature significant deviations from the properties of an equilibrium system~\cite{nikola2016active, cameron2023equation, levis2017active}.
However, if the wall exerts a torque on the gas particles, an equation of state ceases to exist \cite{solon2015pressure}. This has surprising implications. To see this, imagine a hydraulic press where the load may not just depend on pressure differences but also on details of the interactions between the fluid and container walls.

\begin{figure}
	\includegraphics[width=\columnwidth]{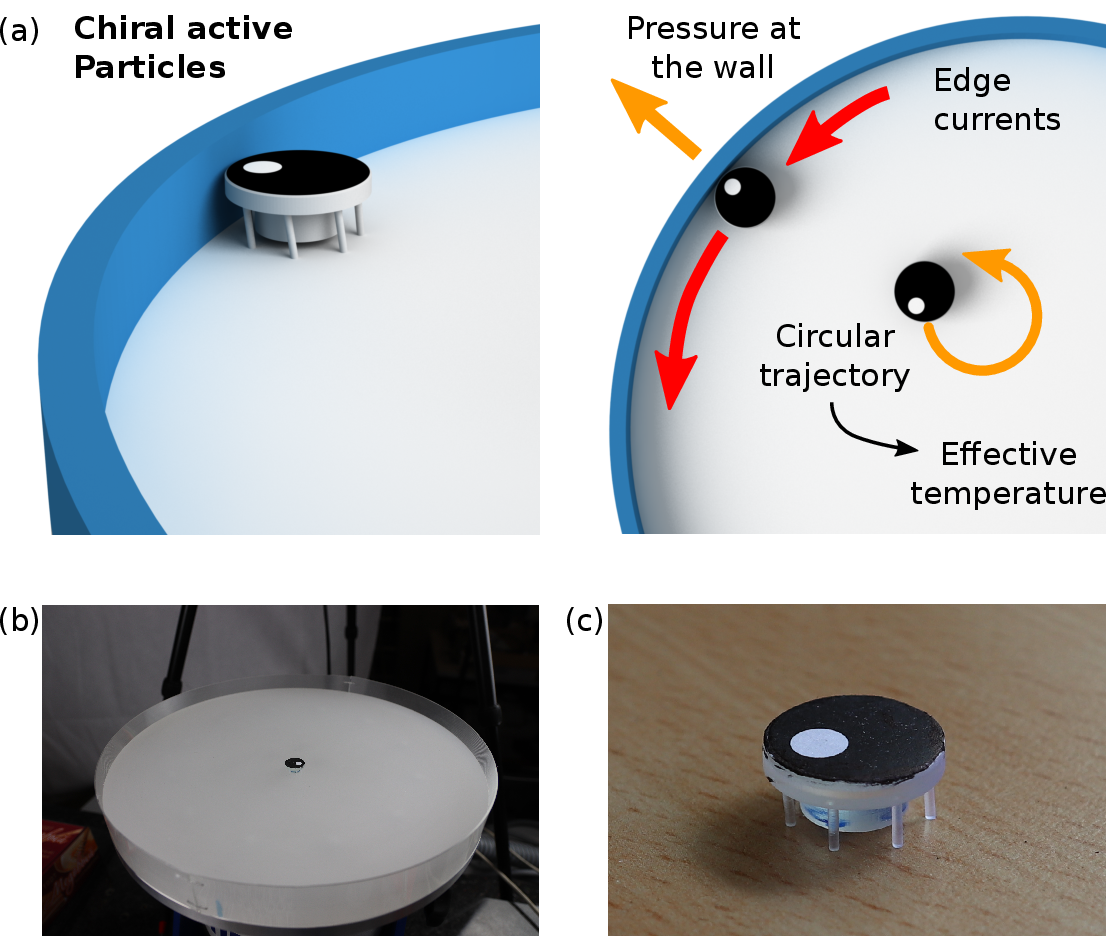}
	\caption{\label{fig:presentation}\textsf{\textbf{Active thermodynamics of a chiral active gas.} 	
(a) Illustration of a chiral active particle realized via a 3D-printed vibrobot which moves on a vibrating plate along noisy circular trajectories. The particle exerts pressure at the container's walls and shows edge currents aligned with chirality, which serve as key ingredients of the active thermodynamics of chiral active gases. (b) Photo of the experimental setup -- a vibrobot on a vibrating plate. (c) Photo of the particle.
	}}
\end{figure}

In contrast to straight-moving active particles, very little is known about the thermodynamic properties of chiral active matter~\cite{lowen2016chirality, liebchen2022chiral}, comprising self-propelling and self-rotating particles which swim in circles or along helices.
This is an important knowledge gap since many, if not most, active systems in the biological world are chiral~\cite{bechinger2016active, woolley2003motility, shi2020chiral}, as a natural consequence of their typical shape asymmetries. Likewise, chirality characterizes also synthetic active systems~\cite{massana2021arrested, lenz2003membranes}, including L-shaped colloidal microswimmers~\cite{kummel2013circular} and isotropic droplets that swim in circles~\cite{carenza2019rotation, sevc2012geometrical}.
For these systems, chirality disrupts wall accumulation typical in active systems~\cite{caprini2019active, van2025three} and alters long-time diffusion coefficients~\cite{van2008dynamics, sevilla2016diffusion, li2020diffusion, khatri2022diffusion, van2022role}. On a collective scale, chirality influences collective phenomena~\cite{liao2018clustering, levis2018micro, levis2019simultaneous, reichhardt2019reversibility, lei2023collective, huang2020dynamical, kuroda2023microscopic, negi2023geometry, debets2023glassy}, inducing clustering inhibition~\cite{ma2022dynamical, kreienkamp2022clustering}, microphase separation~\cite{semwal2024macro}, and vortices~\cite{shee2024emergent, marconi2025spontaneous}, as well as generating angular momentum~\cite{furthauer2012active, marconi2025spontaneous} and self-reverting vorticity in crystals~\cite{caprini2024self} or dilute systems of rotors~\cite{lopez2022chirality}. Recent investigations have delved into the effective chiral dynamics of passive objects immersed in chiral active baths~\cite{poggioli2023odd, hargus2025passive, kalz2025reversal}. This exploration has linked chirality to emergent properties, including odd diffusivity~\cite{hargus2021odd, vega2022diffusive, kalz2022collisions, yasuda2022time,  kalz2024oscillatory, hargus2024flux} -- the spontaneous emergence of orthogonal fluxes relative to motion direction -- experimentally measured in a system of disk-shaped rotors~\cite{vega2022diffusive} and edge currents, experimentally observed in a triangular lattice of self-spinning dimers~\cite{van2016spatiotemporal} or starfish embryos~\cite{tan2022odd}.

These discoveries inspire the question of how chirality impacts the thermodynamic properties of active matter systems. In particular, \emph{a fundamental open question is if chirality alters the thermodynamic properties of an active (self-propelled) ideal} gas and introduces phenomena beyond equilibrium thermodynamics. 

In the present work, we combine theory and simulations with experiments involving granular vibrobots to systematically explore the thermodynamics of inertial chiral active gases, consisting of non-interacting units that self-propel in circular orbits and are characterized by a polarization. As our main results, we derive an equation of state connecting pressure and density via a chirality-dependent effective temperature, as well as a generalization of Fick's law, characterized by chirality-induced off-diagonal diffusion matrix elements producing odd diffusivity and edge currents near walls.
This means that, unlike the achiral active ideal gas, the macroscopic state of a chiral active gas features persistent macroscopic currents (Fig.~\ref{fig:presentation}~(a)) that can be analytically predicted, and show a non-monotonic dependence on chirality due to the particle's inertia.
These results demonstrate that chiral self-propelled gases display activity-induced phenomena even at the ideal gas limit -- ushering in unique thermodynamic concepts like edge currents and odd diffusivity as integral components.

\begin{figure}
	\includegraphics[width=\columnwidth]{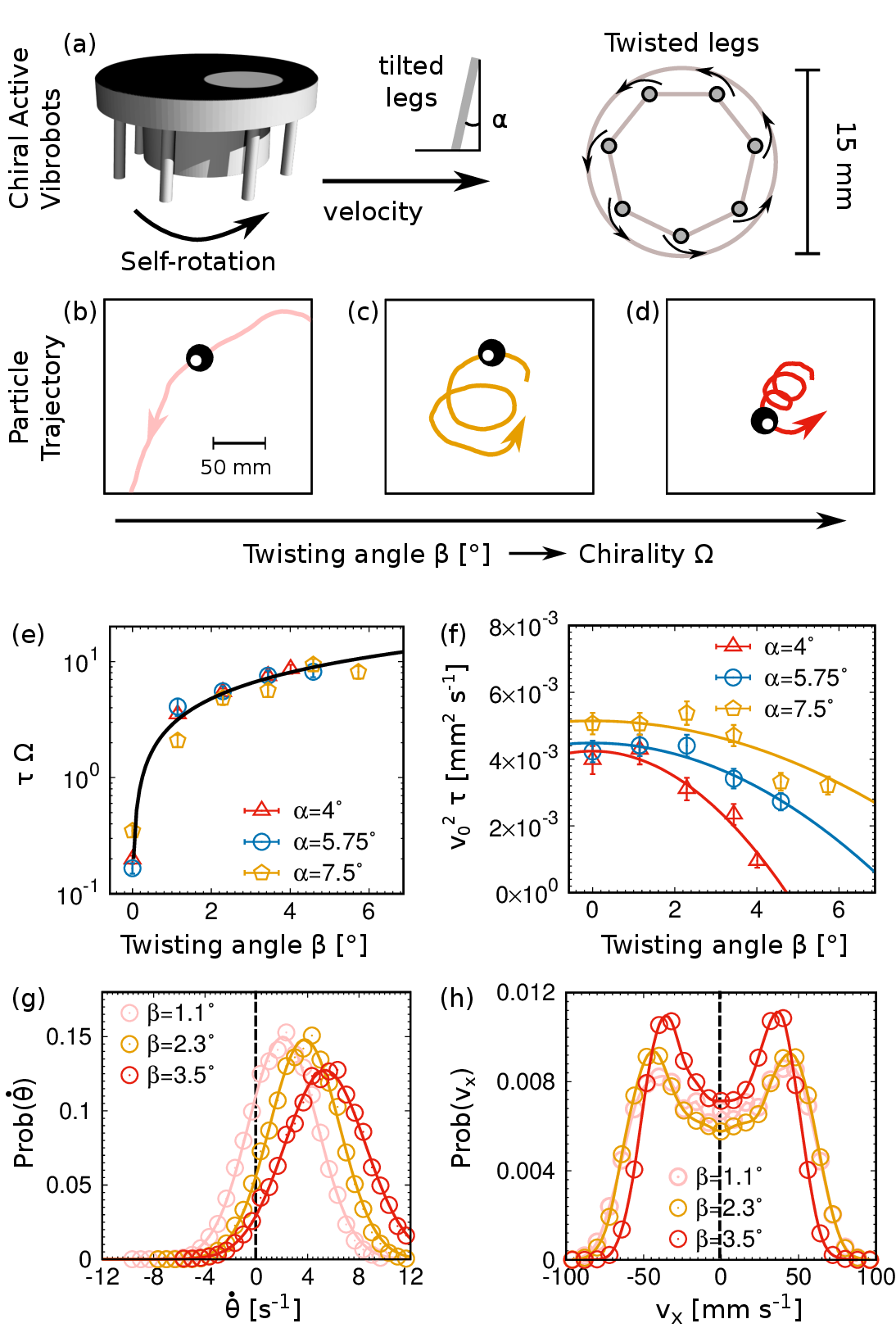}
	\caption{\label{fig:singleparticle}\textsf{\textbf{Chiral active granular particles.} 	
	(a) Illustration of a chiral active granular particle, where legs are tilted by an angle $\alpha$ (tilting angle) and rotated counterclockwise by an angle $\beta$ (twisting angle). $\alpha$ determines the direction of the active velocity, $\beta$ generates self-rotation, while together they induce circular (chiral) motion.
(b)-(d) Experimental time-trajectories ($20$~s) in the plane of motion for $\alpha= 5.75^{\circ}$ and $\beta=1.1^{\circ}, 2.3^{\circ}, 3.5^{\circ}$ in (b), (c), (d).
(e)-(f) Reduced chirality $\Omega\tau$ and active diffusion $v_0^2\tau$ as a function of $\beta$ for several values of $\alpha$. Errors in (e)-(f) are calculated from the standard deviation.
(g)-(h) Probability distributions of angular velocity $\text{Prob}(\dot\theta)$ (g) -- calculated from the instantaneous values of $\dot{\theta}$ -- and translational velocity $\text{Prob}(v_x)$ (h) for different $\beta$ and $\alpha= 5.75^{\circ}$.
In (e)-(h), points are obtained by experiments, while solid lines are guides for the eyes.
	}}
\end{figure}

\paragraph{Experimental setup: a chiral active vibrobot --} 
An ideal gas can be studied by considering many realizations of a single particle experiment.
The particle considered is a chiral active vibrobot, i.e.\ an active granular particle~\cite{kumar2014flocking, baconnier2022selective, deseigne2010collective, kudrolli2010concentration, koumakis2016mechanism, scholz2018inertial, antonov2024inertial} consisting of a cylindrical core with several legs attached (Fig.~\ref{fig:presentation}~(c)). The legs touch a vibrating plate activated by an electromagnetic shaker and surrounded by a plastic confining wall (Fig.~\ref{fig:presentation}~(b)).
By tilting the vibrobot legs by an angle $\alpha$ in the same direction (Fig.~\ref{fig:singleparticle}~(a) and End Matter), the translational symmetry of the particle is broken and the asymmetric bouncing events between legs and plate surface induce directed motion~\cite{caprini2024emergent}. This occurs with an almost-constant self-propulsion speed $v_0$ which is consistent with the double-peaked velocity distribution $\text{Prob}(v_x)$ experimentally measured (Fig.~\ref{fig:singleparticle}~(h)) and theoretically predicted for an active Brownian particle.
The direction of motion, identified by a white eye on the particle top, diffusively changes over time because of imperfections on the plate and legs~\cite{scholz2018inertial}.
By twisting the legs clockwise (or anticlockwise) with the same angle $\beta$ (Fig.~\ref{fig:singleparticle}~(a)), the particle performs self-rotations as a result of a rotational symmetry breaking, as in granular spinners~\cite{scholz2018rotating, farhadi2018dynamics} or circle walkers~\cite{siebers2023exploiting}. This rotational asymmetry induces an angular velocity distribution $\text{Prob}(\dot\theta)$ with a non-zero average $\langle \dot{\theta}\rangle=\Omega>0$ which can be identified with the particle chirality (Fig.~\ref{fig:singleparticle}~(g)).
By tuning the particle design, in particular the tilting ($\alpha$) and twisting ($\beta$) angles of the vibrobot legs, we can control the particle properties, specifically the chirality $\Omega$.
Notably, $\Omega$ exhibits a monotonic increase as a function of $\beta$ and remains independent of $\alpha$ (Fig.~\ref{fig:singleparticle}~(e)).
By contrast, the speed $v_0$ -- extracted from the distribution of the velocity modulus -- monotonically increases with $\alpha$ while it slightly decreases with $\beta$ (Fig.~\ref{fig:singleparticle}~(f)).
The combination of these two effects generates circular trajectories with a radius that increases with the self-propulsion speed - through the tilting angle $\alpha$ - and decreases as the chirality is increased - through the twisting angle $\beta$ - (Fig.~\ref{fig:singleparticle}~(b)-(d)), as expected for a chiral active particle (See Supplementary Movie 1).

\paragraph{Model --}
A previous experimental study~\cite{scholz2018inertial} suggests that our vibrobots behave as chiral active Brownian particles (see SM) described by the following dynamics 
\begin{subequations}
\label{eq:dynamics}
\begin{align}
&m\dot{\mathbf{v}} = -\gamma  \mathbf{v}
+  \,\gamma v_0\boldsymbol{n} + \gamma \sqrt{2 D_t} \,\boldsymbol{\xi} + \mathbf{F}^{w}
 \label{Cdynamicequation2}\\
&\dot{\theta} = \Omega + \sqrt{2D_r}\chi \,,
\label{Cdynamicequation3}
\end{align}
\end{subequations}
where $m$ is the vibrobot mass and $\gamma$ represents the friction coefficient generated by the contact between legs and plate. The motion induced by the asymmetric vertical bouncing of the vibrobot is modeled as an active force $\gamma v_0\mathbf{n}$ where $\mathbf{n}=(\cos{\theta}, \sin{\theta})$ is a unit vector and $\theta$ is the orientational angle of the particle.
The terms $\boldsymbol{\xi}$ and $\chi$ are Gaussian white noises with zero average and unit variance. The first is due to random imperfections on the plate and is responsible for the translational diffusion coefficient $D_t$. The second accounts for random changes in the orientational angle $\theta$, whose strength is determined by the rotational diffusion coefficient $D_r$. Its inverse $\tau=1/D_r$ can be identified with the persistence time, i.e.\ the particle reorientation time. 
The term $\Omega$ in the $\theta$ evolution represents the rotational drift due to the chiral shape of the particle and is often called chirality.
As in previous studies~\cite{caprini2024dynamical}, translational inertia cannot be neglected because the inertial time $m/\gamma$ is only one order of magnitude smaller than the persistence $\tau$: As a consequence, vibrobots are characterized by a real momentum.
A detailed comparison between experiments and simulations is reported in the SM, after extracting the model parameters through an iterative fitting method~\cite{caprini2024dynamical} based on the Nelder-Mead algorithm (see End Matter).
Finally, $\mathbf{F}^w = - \mathbf{e}_r \partial_r V(r)$ is a force term due to a potential $V(r)$ modeling the repulsion generated by the confining walls and pointing along the unit radial vector $\mathbf{e}_r$.

\begin{figure}
	\includegraphics[width=\columnwidth]{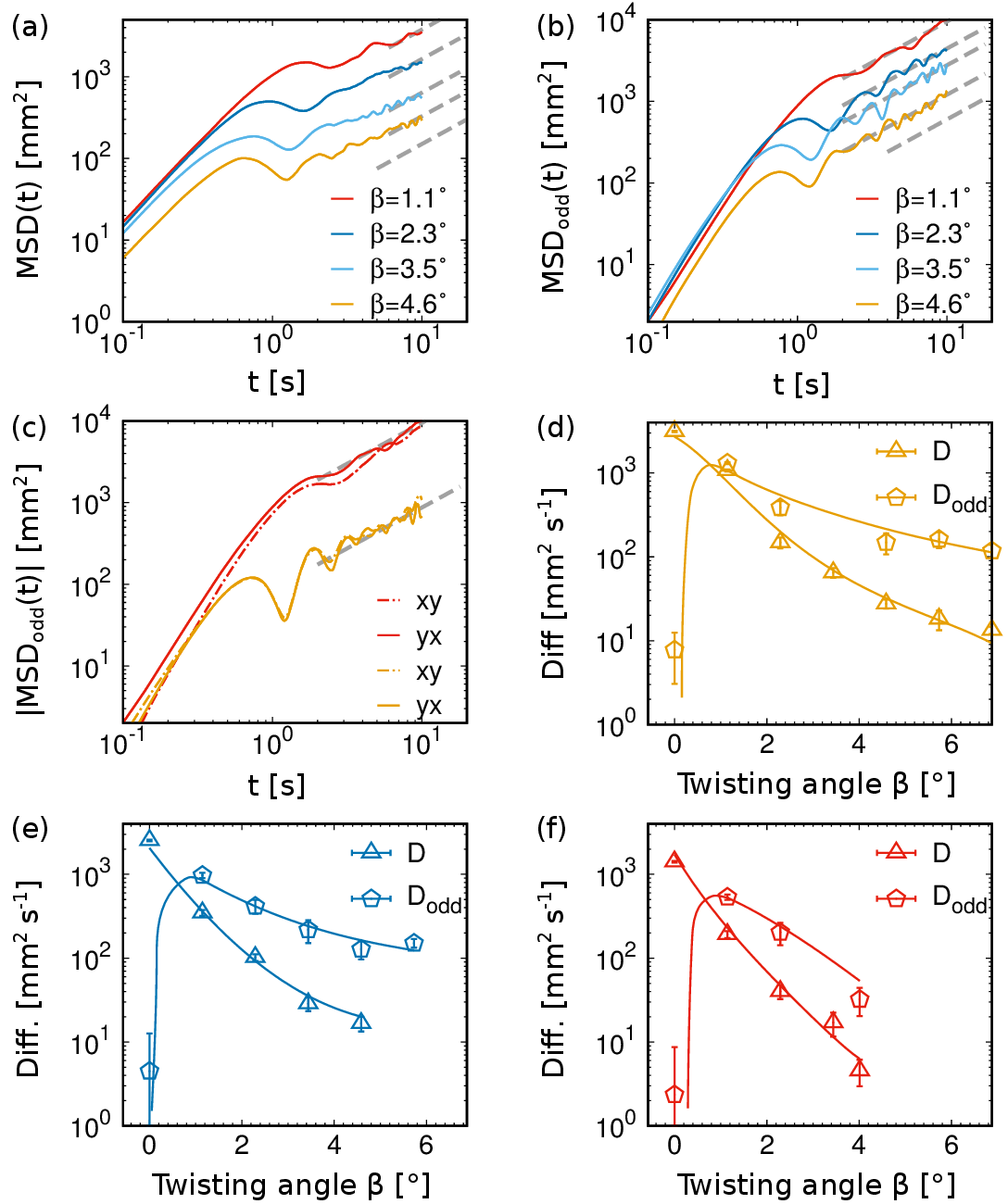}
	\caption{\label{fig:odddiffusion}\textsf{\textbf{Chirality-induced odd diffusion. } 
	(a)-(b)-(c) Mean-square displacements (lines) for several values of the twisting angle $\beta$ and for a tilting angle $\alpha=5.75^{\circ}$. (a) plots the diagonal mean-square displacement, $\text{MSD}(t)$, (b) reports the correlation $\text{MSD}^{xy}_{odd}(t)$, while (c) shows $|\text{MSD}^{xy}_{odd}(t)|$ (dashed-dotted) and $|\text{MSD}^{yx}_{odd}(t)|$ (solid) for $\beta=1.1^{\circ}, 2.3^{\circ}, 3.5^{\circ},  4.6^{\circ}$.
Colored dashed lines in (a)-(c) mark the linear scaling.
(d)-(e)-(f) Long-time diffusion coefficient, $D$, and long-time odd diffusion coefficient, $D_{\text{odd}}$, as a function of $\beta$ for $\alpha= 7.5^{\circ}$ (d), $5.75^{\circ}$ (e), and $4^{\circ}$ (f).
In (d)-(f), points are obtained by experiments and solid colored lines by theoretical predictions, i.e.\ interpolating the values obtained by calculating Eqs.~\eqref{eq:long_time_diff}~\eqref{eq:long_time_diff_odd} with the experimental parameters (see End Matter and SM).
Errors in (d)-(f) are calculated from the standard deviation.
	}}
\end{figure}

\paragraph{Chirality-induced odd diffusion --}
Chirality induces a unique phenomenon in the particle's dynamics known as odd diffusion. This manifests as the particle's ability to diffusively explore the space perpendicular to the displacement direction, through orthogonal fluxes. This implies that the long-time diffusion matrix  $\boldsymbol{\mathcal{D}}$ governing the effective Fick's equation, $\partial_t \rho = \nabla\cdot\boldsymbol{\mathcal{D}}\cdot\nabla \rho$, for the density field $\rho$ is antisymmetric and structured as~\cite{kalz2022collisions}
\begin{equation}
\label{eq:matrix}
\boldsymbol{\mathcal{D}} =
\begin{pmatrix}
D & -D_{odd} \\
D_{odd} & D
\end{pmatrix}\,.
\end{equation}
Here, $D$ represents the long-time diffusion coefficient which can be extracted from the mean-square displacement, $\text{MSD}(t)=\langle (\mathbf{x}(t)-\mathbf{x}(0))^2\rangle$, which displays a ballistic-like regime for a small time, as usual in active systems, and approaches a diffusive-like behavior over long time (Fig.~\ref{fig:odddiffusion}~(a)). As documented in existing literature~\cite{reichhardt2019active, caprini2019active}, chirality introduces oscillations in the $\text{MSD}(t)$ and diminishes $D$ compared to an achiral active particle~\cite{van2008dynamics}. In the inertial chiral active case, $D$ reads
\begin{equation}
\label{eq:long_time_diff}
D = D_t+ \frac{v_0^2 \tau}{1+\Omega^2\tau^2} \,,
\end{equation}
as predicted in the End Matter and verified by comparing Eq.~\eqref{eq:long_time_diff} and the long-time slope extracted from the mean-square displacement (Fig.~\ref{fig:odddiffusion}~(a)).

The term $D_{odd}$, the so-called odd diffusion coefficient, is experimentally extracted by analyzing the long-time behavior of the cross-correlation $\text{MSD}^{xy}_{odd}(t)=t\langle (x(t)-x(0)) v_y(0)\rangle$~\cite{hargus2021odd}. This quantity is accompanied by time oscillations and exhibits a transition from a short-time ballistic behavior to a long-time diffusive regime (Fig.~\ref{fig:odddiffusion}~(b)). The system oddness manifests already in the odd profile of $\text{MSD}^{xy}_{odd}(t)$, such that $\text{MSD}^{xy}_{odd}(t) = - \text{MSD}^{yx}_{odd}(t)$ (Fig.~\ref{fig:odddiffusion}~(c)).
For a single chiral active granular particle, characterized by inertia, the odd diffusion coefficient can be predicted analytically (see SM):
\begin{equation}
\label{eq:long_time_diff_odd}
D_{odd} = v^2_0 \tau  \frac{\Omega \tau}{1+\Omega^2\tau^2} \frac{1+\frac{2m}{\gamma\tau}}{\left( 1+ \frac{m}{\gamma \tau} \right)^2 + \frac{m^2 \Omega^2}{\gamma^2}}  \,.
\end{equation}
Interestingly, the chirality $\Omega$ induces a non-monotonic behavior in $D_{odd}$. Indeed, for vanishing chirality $D_{odd}=0$, the particle shows only standard diffusion. 
By contrast, a larger $\Omega$, induced by an increasing twisting angle $\beta$, hinders the particle's ability to diffusively explore the surrounding space both translationally and orthogonally to the direction of motion.
This effect quantitatively predicted by Eq.~\eqref{eq:long_time_diff_odd} is investigated for different values of the tilting angle $\alpha$ (Fig.~\ref{fig:odddiffusion}~(d)-(e)-(f)).
Experimentally, the case $\beta=0$ has a residual chirality which results in an odd diffusion $D_{odd}$ almost two orders of magnitude smaller than $D_{odd}$ corresponding to the first non-vanishing $\beta$ value experimentally accessible. From further $\beta$ values, $D_{odd}$ starts decreasing in agreement with our chirality estimate, such that $\Omega \tau>1$.
As a consequence, odd diffusivity is maximized by the interplay of two time scales, namely, the persistence time $\tau$ and the revolution time $1/\Omega$. Remarkably, compared to magnetic systems, our odd diffusion coefficient can reach values almost one order of magnitude larger than standard diffusion.

\paragraph{Equation of state --}

\begin{figure}
	\includegraphics[width=\columnwidth]{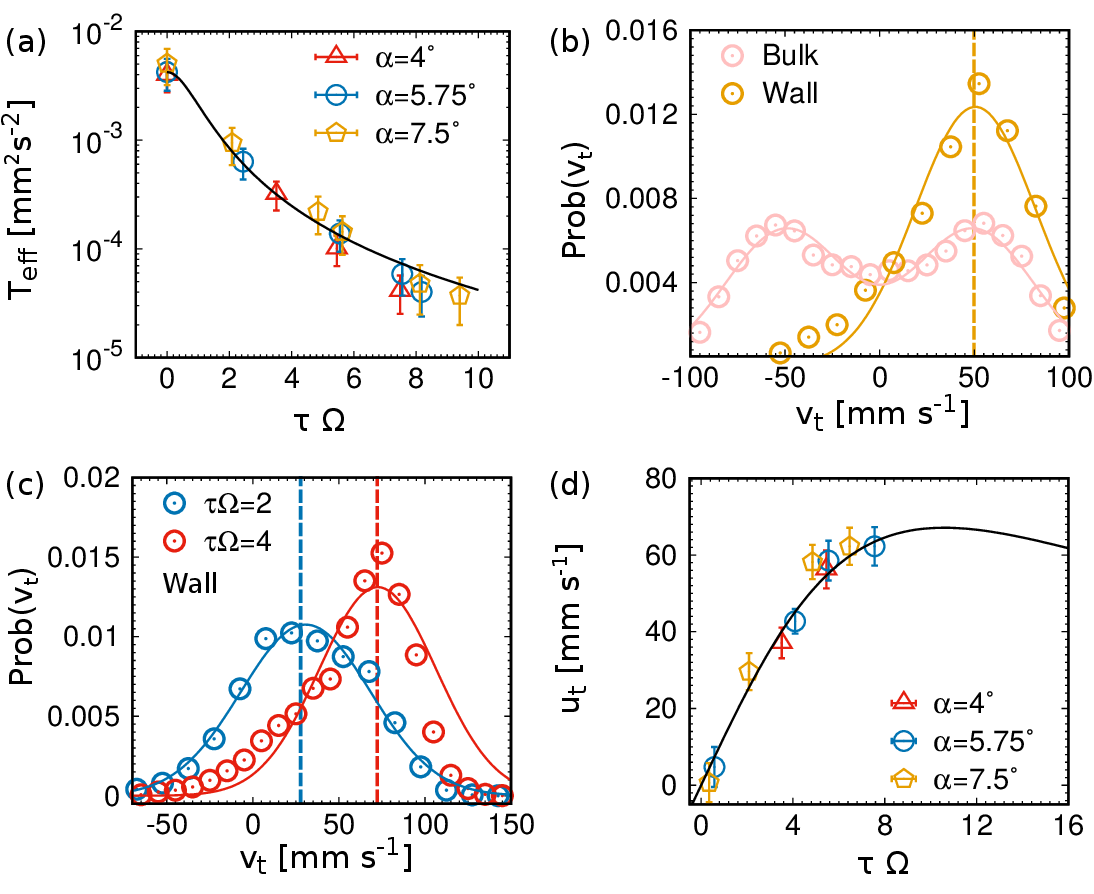}
	\caption{\label{fig:momentum}\textsf{\textbf{Effective temperature and edge currents.} 
	(a) Effective temperature $T_{eff} =\gamma D_t + \frac{v_0^2\tau\gamma}{1+\Omega^2\tau^2}$ as a function of the rescaled chirality $\tau\Omega$ - generated by different twisting angles $\beta$ and tilting angles $\alpha$.
(b) Probability distribution $\text{Prob}(v_t)$ of the tangential velocity $v_t$ from the middle of the plate, for particles moving close (wall) and far (bulk) from the wall for $\tau\Omega=3$.
(c) $\text{Prob}(v_t)$ close to the wall for two values of $\tau\Omega$.
(d) Average tangential velocity $u_t=\langle v_t(\mathbf{x}\approx R \rangle$ calculated close to the wall. 
Colored points are obtained from experimental data, black lines in (a) and (d) represent the scaling functions $\sim1/(1+\tau^2\Omega^2)$ and $\sim \Omega\tau/(1+ k\,\tau^2 \Omega^2)$, while colored curves in (b) and (c) are guides for the eyes.
Errors in (a) and (d) are calculated from the standard deviation.
	}}
\end{figure}

To derive the equation of state, we resort to a hydrodynamic theory for the slow fields governing the dynamics: the density $\rho(\mathbf{x})$, the velocity field $\mathbf{u}(\mathbf{x})=\langle \mathbf{v}(\mathbf{x})\rangle/\rho$ and the polarization field $\mathbf{p}(\mathbf{x}) = \langle \mathbf{n}(\mathbf{x})\rangle$ -- see End Matter for the fields' definition. While $\rho$ is related to $\mathbf{u}$ through the continuity equation, the momentum equation is linearly coupled to the polarization - see the End Matter for the exact closure for our hydrodynamic theory. The dynamics for $\mathbf{p}$ is determined by the chirality and contains orthogonal fluxes coupling $x$ and $y$ components of $\mathbf{p}$ as an effect of odd diffusion.
By integrating the momentum equation from the bulk to the wall region, it is possible to exactly calculate the mechanical pressure $\mathcal{P}_m$. The circular symmetry of the potential suggests the use of polar coordinates, for which the pressure exerted on the wall is defined as $\mathcal{P}_m = \int_{0}^{\infty} dr \int_0^{2\pi} d\phi \dens(r)\, \partial_r V(r)$. 
With this method, we discover that a gas of inertial chiral active particles is governed by an equation of state, relating the mechanical pressure and the density (see End Matter):
\begin{equation}
\label{eq:Pressure_pred}
\mathcal{P}_m 
=\left( \gamma D_t + \frac{v_0^2\tau\gamma}{1+\Omega^2\tau^2} \right)  \rho_0  \,,
\end{equation} 
where $\rho_0$ is the radial density value in the bulk, e.g.\ far from the wall.
The term in brackets can be identified as the effective temperature of the system, $T_{\text{eff}}=\gamma D_t + \frac{v_0^2\tau\gamma}{1+\Omega^2\tau^2}$, which consists of a thermal contribution analog to equilibrium and an active one $\propto \gamma v_0^2\tau$. The latter term decreases when the chirality $\tau\Omega$ is increased - through the twisting angle $\beta$ - as verified for different values of the tilting angle $\alpha$ (Fig.~\ref{fig:momentum}~(a)).
Indeed, the circular trajectories induced by chirality reduce the particle's ability to explore the surrounding space and thus the so-called swim pressure~\cite{takatori2014swim, takatori2015towards, takatori2016forces} responsible for the increase of the mechanical pressure in active systems.

\paragraph{Edge currents --}
By evaluating the momentum equation in the direction tangential to the wall profile, we are able to analytically predict the occurrence of edge currents parallel to the wall. In the steady state, the velocity field component orthogonal to the wall profile vanishes $u_r=0$, and the density gradient along the radial direction is related to the wall force (see End Matter).
As a result, we obtain a non-vanishing steady-state velocity field parallel to the wall $u_t$
\begin{equation}
\label{eq:ut_prediction}
\gamma\dens  u_t(r)= 
 \frac{D_{odd}}{D} \dens(r) \,\partial_r V(r) \,.
\end{equation}
This result $\propto \dens\mathcal{D}_{odd}/D$ predicts edge currents aligned with chirality $\Omega$: as expected, this term changes sign when the $\Omega$ is reversed and vanishes when the wall is absent ($V(r)=0$) or $\Omega=0$.

Edge currents are experimentally measured by calculating the tangential component of the vibrobot velocity $v_t$ from the middle of the circular plate. The probability distribution $\text{Prob}(v_t)$ is fully symmetric for a particle moving far from the wall, while $\text{Prob}(v_t)$ is peaked at $v_t > 0$ when the particle moves close to the container boundary (Fig.~\ref{fig:momentum}~(b)). This peak shifts to the right as $\tau\Omega$ increases (Fig.~\ref{fig:momentum}~(c)): This implies that the tangential component of the velocity field, $u_t(r\approx R)$ -- at the wall coordinate $r\approx R$ -- does not vanish.
Our experiments confirm Eq.~\eqref{eq:ut_prediction} because $u_t(r\approx R)$ increases with $\tau\Omega$ -- controlled by the twisting angle $\beta$ -- as checked for different tilting angles $\alpha$ (Fig.~\ref{fig:momentum}~(d)). This is in agreement with the scaling $u_t \sim D_{\text{odd}}/D \sim \tau\Omega/(1+k\,\tau^2\Omega^2)$ theoretically predicted, where $k\approx m^2/(\gamma^2\tau^2)$ (see SM) is the inertial correction responsible for the deviation from the linear increase and the occurrence of a non-monotonic behavior when $m/\gamma$ is increased.
We remark that $D_{odd}$ and $D$ appearing in Eqs.~\eqref{eq:ut_prediction}-\eqref{eq:Pressure_pred} are bulk observables, while their values at the boundary could be affected by $u_t$.

Our hydrodynamic theory is capable of reproducing experimental data for an ideal chiral active gas consisting of chiral active granular particles, because of the small Knudsen number value~\cite{chapman1990mathematical} experimentally measured, $\mathcal{K}_\tau= v_0 \tau/(2R) \ll 1$, defined as the ration between the persistence length $v_0\tau$ and the plate diameter $2R$ (see SM).
This theory holds even if we have neglected non-linear terms because these additional contributions generate higher-order gradients which do not affect the Fick-law governing non-interacting particles.
The generality of our theoretical approach is limited to ideal gases consisting of self-propelled rotating units, i.e.\ non-interacting chiral active particles characterized by a polarization vector.
This implies that our theory would need to be generalized before describing non-polar granular spinners, where the effective temperature has been previously explored~\cite{farhadi2018dynamics, han2021fluctuating}, and the particles transfer angular momentum during interactions~\cite{caporusso2024phase, caprini2024bubble, digregorio2025phase}.

\paragraph{Conclusions --}
The present work shows that edge currents and odd diffusivity represent the peculiar thermodynamic features of an ideal chiral active gas with inertia.
Even if the system is far from equilibrium, these properties and the pressure exerted on the container's wall can be related to bulk properties, such as the effective temperature and the odd diffusivity.
We have provided explicit predictions for these observables, which can be used in the future to design the oddity of active systems. This might be interesting for applications; for instance, edge currents could be used to clean up boundaries from debris or to increase the efficiency of motors powered by active particles~\cite{di2010bacterial, hiratsuka2006microrotary}.
Our experimental chiral active vibrobots are good candidates to experimentally observe odd viscosity~\cite{fruchart2023odd, banerjee2017odd, lou2022odd, markovich2021odd, tan2022odd, ding2024odd} and odd elasticity~\cite{scheibner2020odd, braverman2021topological, surowka2023odd, kobayashi2023odd}, which have been explored by macroscopic theories.

\paragraph{Acknowledgments --} 
LC acknowledges financial support from the University of Rome Sapienza, under the project Ateneo 2024 (RM124190C54BE48D). 
HL acknowledges support by the Deutsche Forschungsgemeinschaft (DFG) through the SPP 2265 under the grant number LO 418/25.


\bibliographystyle{apsrev4-2}
\bibliography{bib}

\appendix

\section{Experimental details}

\noindent
\textbf{Experimental setup --}
To induce vibrational excitations in granular materials, we employ a standard vibrating table. 
Our setup comprises an acrylic baseplate with a diameter of $\unit[300]{mm}$ and a height of $\unit[15]{mm}$, complemented by an outer plastic ring that confines the particles within.
Vertical vibrations are generated by an electromagnetic shaker which is affixed to the acrylic baseplate, aligned in the horizontal direction to suppress any gravitational drift. Additionally, the setup is placed on a substantial concrete block to counteract resonance effects with the environment.
The chosen shaker frequency $f=\unit[120]{Hz}$ and amplitude of $A=\unit[24(1)]{\upmu m}$ ensures quasi-two-dimensional particle motion on the baseplate.

\vskip4pt
\noindent
\textbf{Tracking --}
We employ a high-speed camera, recording the system at a rate of $50$ frames per second with a resolution of $1024\times1024$ pixels. To determine the positions, $\mathbf{x}$, and orientations, $\theta(t)$, of the vibrobots, a standard feature recognition method is utilized - specifically the circle Hough transform complemented by a custom classical algorithm for sub-pixel localization.
The reconstruction of particle trajectories involves tracking nearest neighbors between the current frame and a pseudo-frame, extrapolated from the known positions and velocities of the previous frame. Translational and angular velocities $\mathbf{v}(t)$ and $\dot{\theta}(t)$ are calculated from the position and angle displacements using the centered finite difference formulas, such that: $\mathbf{v}(t)=(\mathbf{x}(t+\Delta t)-\mathbf{x}(t-\Delta t))/(2\Delta t)$ and $\dot{\theta}(t)=(\mathbf{\theta}(t+\Delta t)-\mathbf{\theta}(t-\Delta t))/(2\Delta t)$, with $\Delta t=\unit[0.02]{s}$. Given the relatively large inertial relaxation time compared to the data acquisition time resolution, these displacements provide an accurate approximation of the particles' instantaneous translational and angular velocities.
The probability distributions $\text{Prob}(\dot{\theta})$ and $\text{Prob}(v_x)$ are calculated from the instantaneous values of $\dot{\theta}(t)$ and $v_x(t)$ considered at independent times, i.e. after a time interval of $1$~s comparable with the value of the persistence time $\tau=1/D_r$.

\vskip4pt
\noindent
\textbf{Active chiral vibrobot design --}
The particles are three-dimensional plastic objects created through 3D printing~\cite{scholz2018inertial} with a mass of $m=\unit[0.83]{g}$.
The particle body comprises three primary components:
i) A cylindrical core with a diameter of $\unit[9]{mm}$ and a height of $\unit[4]{mm}$ to stabilize the particle by lowering its center of mass vertically.
ii) A larger cylindrical cap, with a diameter of $\unit[15]{mm}$ and a height of $\unit[2]{mm}$, which defines the particle's circular horizontal cross-section.
iii) Seven cylindrical legs attached to the cap, each with a diameter of $\unit[0.8]{mm}$ and a vertical height of $\unit[5]{mm}$. These legs are regularly distributed around the core, forming an eptagonal configuration. Since the legs are in contact with the ground, the particle has a total height of $\unit[7]{mm}$.

To induce self-propelled motion, we introduce a disruption in the particle's translational symmetry by tilting all the legs in the same direction with an angle $\alpha$  relative to the surface normal. The tilting direction is indicated by a white spot attached to the black particle's cap.
We experimentally consider $\alpha=4^{\circ}, 5.75^{\circ}, 7.5^{\circ}$: the larger $\alpha$, the higher the particle velocity.
Self-rotational motion is achieved by breaking the cylindrical symmetry of the particle: all the legs are twisted counterclockwise, i.e.\ with another angle relative to the vertical $\beta$. Experiments are realized with $\beta=1.1^{\circ}, 2.3^{\circ}, 3.4^{\circ}, 4.6^{\circ}, 5.7^{\circ}$:
The larger $\beta$, the higher the particle's angular velocity.
Circular motion is accomplished by simultaneously tilting and twisting the legs, combining translational and rotational symmetry-breaking mechanisms.

\vskip4pt
\noindent
{\bf Parameter extraction --}
The values of the parameters $v_0$, $D_r$ (or $\tau=1/D_r$), $D$, $\gamma$, and $\Omega$ are extracted by performing simulations of Eqs.~\eqref{eq:dynamics} through a Nelder-Mead optimization scheme~\cite{nelder1965simplex}. This is a derivative-free optimization method for finding a local minimum of a function of several variables, obtained by iteratively replacing the parameters' values until a cost function is minimized. This cost function is calculated as the square difference between observables measured in experiments and simulations, including the translational and angular velocity distributions, the mean square displacement, and the angular mean-square displacement. For further details on the parameter estimate and the comparison between experiments and simulation, see SM.

\section{Coarse-grained theory}
A hydrodynamic description can be formulated by projecting the Fokker-Planck equation (FPE) for an inertial chiral active particle for the probability distribution $f=f(\mathbf{x}, \mathbf{v}, \mathbf{n})$ on the velocity and self-propulsion momenta. 
In particular, by integrating the FPE over $\mathbf{n}$ and $\mathbf{v}$, we obtain the continuity equation for the density field $\rho=\rho(\mathbf{x})=\int d\vv\, d\hn \, f$; by multiplying the FPE by $\mathbf{n}$ and $\mathbf{v}$ and integrating, we can derive the balance equation for the polarization $\mathbf{p}=\mathbf{p}(\mathbf{x})=\int d\vv\, d\hn \, f \mathbf{n}$ and the velocity fields $\mathbf{u}=\mathbf{u}(\mathbf{x}) = \int d\vv\, d\hn \, f \mathbf{v}$: 
\begin{subequations}
\label{eq:hydro_app}
\begin{align}
&
\partial_{t}\rho = - \nabla\cdot(\rho\, {\bf u}) \label{eq:cont}\\
&
\partial_{t}{\bf u} + {\bf u}\cdot\nabla{\bf u} = \frac{1}{\rho}\frac{\gamma}{m} v_0 {\bf p} - \frac{1}{m \rho}\nabla\cdot {\bf P} - \frac{\gamma}{m}{\bf u} + \frac{\mathbf{F}_{ex}}{m}
\label{eq:mom}\\ 
&
\partial_{t} {\bf p} + \nabla\cdot  (\rho \,\frac{\boldsymbol{\mathcal{W}}}{m v_0})=- \frac{\boldsymbol{\mathcal{A}}}{\tau} \cdot \bf{p} \,,\label{eq:pol}
\end{align}
\end{subequations}
where $\boldsymbol{\mathcal{A}}$ is an antisymmetric matrix with elements $\mathcal{A}_{ij}=\delta_{ij} + \Omega \tau\, \epsilon_{ij}$ with $\epsilon_{ij}$ the two-dimensional Levi Civita tensor, such that $\epsilon_{12}=1$ and $\epsilon_{21}=-1$.
Here, the symbol ${\bf P}$ denotes the pressure tensor due to other particles or kinetic contribution while the tensor $\boldsymbol{\mathcal{W}}=m v_0 \langle \hn\otimes\mathbf{v} \rangle$ may be interpreted as the generalized work performed by each component of the activity.
Compared to the passive case, the evolution for the velocity field *Eq.~\eqref{eq:mom}) non-reciprocally couples to the polarization equation, via an explicit linear term.
In addition, Eq.~\eqref{eq:pol} is determined by the work performed by the active force and, compared to the hydrodynamic theory of linear inertial active particles~\cite{epstein2019statistical, steffenoni2017microscopic}, this equation antisymmetrically couples different spatial components of the polarization via the chirality term contained in the antisymmetric matrix $\boldsymbol{\mathcal{A}}$.

The hydrodynamic equations~\eqref{eq:cont}-\eqref{eq:pol} are not close because of the presence of second-order tensors, i.e.\ $\boldsymbol{\mathcal{W}}$ and ${\bf P}$.
Thus, to close the hierarchy, we consider the hydrodynamic equation for the work tensor and the kinetic temperature fields (see SM). 
By neglecting spatial gradients (see SM), we can approximate $\boldsymbol{\mathcal{W}}$ as
\begin{equation}
\label{eq:approxW}
\dens \,\boldsymbol{\mathcal{W}} \approx \dens \boldsymbol{\mathcal{M}}^{-1}   \gamma v_0^2 \,.
\end{equation}
where the matrix $\boldsymbol{\mathcal{M}}^{-1}$ reads
\begin{equation}
\boldsymbol{\mathcal{M}}^{-1} = \frac{\tau}{(1+\tau\frac{\gamma}{m})^2+\tau^2\Omega^2}
\begin{pmatrix}
\label{eq:matrix_Mm1}
1+\tau\frac{\gamma}{m} &\tau\Omega\\
\tau\Omega&1+\tau\frac{\gamma}{m}
\end{pmatrix}\,.
\end{equation}
The presence of chirality generates odd out-of-diagonal components in $\boldsymbol{\mathcal{W}}$, via the matrix $\boldsymbol{\mathcal{M}}^{-1}$, while, for vanishing $\Omega\to0$, the tensor $\boldsymbol{\mathcal{W}}$ has a diagonal structure.
To close the theory, we linearize the pressure tensor ${\bf P}=P \,{\bf I}$, obtaining (see SM)
\begin{equation}
\label{eq:approxP}
P  \approx   \gamma D_t \,\rho +  \text{Trace}[\boldsymbol{\mathcal{W}}] \frac{\rho}{2}\,. 
\end{equation}
By neglecting time derivative in Eq.~\eqref{eq:pol} and using the approximation~\eqref{eq:approxW} and~\eqref{eq:approxP},
we obtain a close expression for the polarization $\bf{p}$ as function of the density $\rho$: 
\begin{equation}
\mathbf{p}=- \frac{1}{m v_0}\tau\boldsymbol{\mathcal{A}}^{-1}\cdot \boldsymbol{\mathcal{W}}\cdot\nabla \dens \,.
\end{equation}
As usual in overdamped active particles, the polarization gives rise to an additional pressure term, often called swim or active pressure~\cite{takatori2014swim, takatori2015towards, takatori2016forces} that explicitly depends on the work tensor, in inertial active particles.
In inertial chiral particles, the polarization is decreased by chirality because of the $\Omega$-dependence in the matrix $\boldsymbol{\mathcal{M}}^{-1}$.
Finally, by neglecting the time dependence in the velocity fields, neglecting non-linear terms, and using the closure for $\mathbf{P}$ (Eqs.~\eqref{eq:approxW} and~\eqref{eq:approxP}), the relation~\eqref{eq:mom} becomes
\begin{equation}
\label{eq:mom_eq_ss}
\dens \frac{\gamma}{m} \mathbf{u} = -\frac{\gamma}{m}  \boldsymbol{\mathcal{D}} \cdot \nabla \dens  +  \frac{\mathbf{F}_{ex}}{m}\rho \,,
\end{equation}
where the matrix $\boldsymbol{\mathcal{D}}$ can be expressed as a function of the work tensor $\boldsymbol{\mathcal{W}}$ as
\begin{equation}
\label{eq:diffusion_fick}
\boldsymbol{\mathcal{D}}={\bf}  \left(D_t + \frac{1}{2} \, \frac{\text{Trace}[\boldsymbol{\mathcal{W}}]}{\gamma}\right){\bf I} + \frac{\tau}{m} \boldsymbol{\mathcal{A}}^{-1}\cdot\boldsymbol{\mathcal{W}} \,.
\end{equation}
The expression~\eqref{eq:diffusion_fick} corresponds to Eq.~\eqref{eq:matrix}. 

\vskip4pt
\noindent
{\bf Unconfined chiral active particles --}
The matrix $\boldsymbol{\mathcal{D}}$ can be identified with the diffusion matrix of the system. 
Indeed, by inserting Eq.~\eqref{eq:mom_eq_ss} with $\mathbf{F}_{ex}=0$ in the continuity equation~\eqref{eq:cont}, we obtain a Fick's equation 
\begin{equation}
\partial_{t}\rho = \nabla \cdot \boldsymbol{\mathcal{D}} \cdot\nabla \rho\,.
\end{equation}
This equation implies that inertial chiral systems are governed by odd diffusion.

\vskip4pt
\noindent
{\bf Circular confinement --}
We chose a radially symmetric external force $\mathbf{F}^{ex}=\mathbf{e}_r F_r=- \mathbf{e}_r \partial_r V(r)$, such that $V(r)=h(r) \Theta(r-R)$ where $\Theta(r-R)$ is the step-function and $h(r)$ represents the wall profile ($h(r=R)=0$) which models the effect of boundaries on the circular plate.
The symmetry of the potential suggests the use of radial and tangential coordinates calculated from the origin, along the unit vectors $\mathbf{e}_r = \mathbf{e}_x \cos\phi + \mathbf{e}_y \sin\phi$ and $\mathbf{e}_t=\mathbf{e}_x \sin\phi - \mathbf{e}_y \cos\phi$, where $\mathbf{e}_x$ and $\mathbf{e}_y$ are the unit vector along the two Cartesian components and $\phi$ is the polar angle.
In polar coordinates, Eq.~\eqref{eq:mom_eq_ss} becomes
\begin{subequations}
\begin{align}
\label{eq:mom_rad_ss_app}
&\frac{\gamma}{m}\rho \,u_r = - \frac{\gamma}{m}  \left(  D  \nabla_r \rho - D_{\text{odd}} \frac{1}{r}\nabla_\phi \rho\right) + \frac{F_r}{m}\rho\\
\label{eq:mom_tan_ss_app}
&\frac{\gamma}{m}\rho\,u_t = - \frac{\gamma}{m} \left( D \frac{1}{r}\nabla_\phi \rho+D_{\text{odd}} \nabla_r \rho \right) \,.
\end{align}
\end{subequations}
The particle is confined along $\mathbf{e}_r$ and thus $u_r=0$, while it is free to move along $\mathbf{e}_t$ where $\nabla_\phi \rho =0$ in the steady-state.

\vskip4pt
\noindent
{\bf Equation of state --}
By introducing $\tilde{\rho}(r) = \int_0^{\pi} \rho(r, \phi)$ and defining the mechanical pressure as 
\begin{equation}
\label{eq:pressure_def}
\mathcal{P}_m = \int_0^{\infty} dr \tilde{\rho}(r) \partial_r V(r) = - \int_0^{2\pi} d\phi \int_0^{\infty}dr \rho F_r\,,
\end{equation}
we obtain the equation of state by replacing the steady-state version of Eq.~\eqref{eq:mom_rad_ss_app} in Eq.~\eqref{eq:pressure_def}:
\begin{equation}
\begin{aligned}
\frac{\mathcal{P}_m}{m}&=
- \frac{\gamma}{m} \int_0^{\infty} dr \int_0^{2\pi} d\phi   D  \nabla_r \rho =\frac{\gamma}{m} D  \left[\tilde{\rho}_0 - \tilde{\rho}_\infty\right] \\
&=\left( \frac{\gamma D_t}{m} + \frac{\gamma}{m}\frac{v_0^2\tau}{1+\Omega^2\tau^2} \right)  \tilde{\rho_0} \,,
\end{aligned}
\end{equation}
where $\tilde{\rho}_0$ is the bulk radial density.

\vskip4pt
\noindent
{\bf Edge currents --} 
The steady-state version of Eq.~\eqref{eq:mom_rad_ss_app} allows us to find an expression for $\nabla_r \rho$
\begin{equation}
\frac{\gamma}{m}D\nabla_r \rho = \frac{F_r}{m}\rho \,.
\end{equation}
By inserting this result in the equation for $u_t$ (Eq.~\eqref{eq:mom_tan_ss_app}), we have
\begin{equation}
\begin{aligned}
&\frac{\gamma}{m} \rho \,u_t =  - \frac{D_{\text{odd}}}{D} \frac{F_r}{m}\rho
\end{aligned}
\end{equation}
which corresponds to Eq.~\eqref{eq:ut_prediction} and predicts edge currents.

\end{document}